\documentclass[openacc]{rstransa}

\usepackage{cite}
\usepackage{amsmath,amssymb,amsfonts}
\usepackage{algorithmic}
\usepackage{graphicx}
\usepackage{textcomp}
\usepackage{url}            % simple URL typesetting
\usepackage{booktabs}       % professional-quality 
\usepackage{nicefrac}       % compact symbols for 1/2, etc.
\usepackage{microtype}      % microtypography
\usepackage{subfig}
\usepackage{booktabs} % for professional tables
\usepackage{bm} 
\usepackage{bbm}
\usepackage{hyperref}
\usepackage{xcolor}
\usepackage{ulem}

\usepackage[numbers]{natbib}

\graphicspath{{./Figures/}}

\definecolor{darkgreen}{rgb}{0,0.5,0}
\definecolor{redred}{HTML}{D53E4F}
\definecolor{darkred}{HTML}{9B0000}
\definecolor{greengreen}{HTML}{1C911E}
\definecolor{oran}{HTML}{D6923C}
\definecolor{yell}{HTML}{D1C72E}
\definecolor{gree}{HTML}{5CB14E}
\definecolor{cycy}{HTML}{4EBCB3}
\definecolor{bluu}{HTML}{5278BD}
\definecolor{mage}{HTML}{A14BA1}
\definecolor{viol}{HTML}{924FA4}

\newcommand{\str}{{\color{redred} r}}
\newcommand{\stg}{{\color{gree} g}}
\newcommand{\stb}{{\color{bluu} b}}

\newcommand{\bea}{\begin{eqnarray}}
\newcommand{\eea}{\end{eqnarray}}

\newcommand{\jvec}{{\bf j}}

\begin{document}

\title{Loop-Free Tensor Networks for High-Energy Physics}

\author{%%%% Author details
Simone Montangero$^{1,2,3}$, Enrique Rico$^{4,5}$, Pietro Silvi$^{6}$ }

%%%%%%%%% Insert author address here
\address{
$^{1}$Dipartimento di Fisica e Astronomia ``G. Galilei'', Universit\`a di Padova, I-35131 Padova, Italy\\
$^{2}$Istituto Nazionale di Fisica Nucleare (INFN), Sezione di Padova, I-35131 Padova, Italy\\
$^{3}$Padua Quantum Technologies Research Center, Universit\`a degli Studi di Padova, I-35131 Padova, Italy\\
$^{4}$Department of Physical Chemistry, University of the Basque Country UPV/EHU, Apartado 644, 48080 Bilbao, Spain \\
$^{5}$IKERBASQUE, Basque Foundation for Science, Plaza Euskadi 5, 48009 Bilbao, Spain\\
$^{6}$Center for Quantum Physics, and Institute for Experimental Physics, University of Innsbruck, A-6020 Innsbruck, Austria\\
}

%%%% Subject entries to be placed here %%%%
\subject{High Energy Physics, Numerical Simulation of Complex Quantum Systems}

%%%% Keyword entries to be placed here %%%%

%%%% Insert corresponding author and its email address}
\corres{Simone Montangero\\
\email{simone.montangero@unipd.it}}

\begin{abstract}
This brief review introduces the reader to tensor network methods, a powerful theoretical and numerical paradigm spawning from condensed matter physics and quantum information science and increasingly exploited in different fields of research, from artificial intelligence to quantum chemistry. Here, we specialise our presentation on the application of loop-free tensor network methods to the study of High-Energy Physics (HEP) problems and, in particular, to the study of lattice gauge theories where tensor networks can be applied in regimes where Monte Carlo methods are hindered by the sign problem.
\end{abstract}

\begin{fmtext}
\vspace{-1cm}
\section{Introduction}

Lattice QCD is the framework to define the non-perturbative dynamics of strong interactions from first principles. Both the algorithmic development and the increase of the performance of the supercomputer over past forty years enabled numerical simulations able to predict physical quantities associated with single hadrons with high precision. The standard approach of lattice gauge theories (LGT) relies on Monte Carlo-based evaluations of path integrals in Euclidean space-time with positive integrands. 
\end{fmtext}

\maketitle

\noindent Despite being extremely successful, it suffers from a fundamental limitation in scenarios that give rise to a sign problem. These scenarios include the presence of a finite baryon density, which is relevant for the early universe and for neutron stars; real-time evolution, e.g., to understand the dynamics of heavy-ion collisions; or topological terms, which could shed light on the matter-anti-matter asymmetry of the universe. 
It has been shown that mitigating the sign problem is NP-hard \cite{SignProblemNPHard}. There is therefore an urgent quest to find alternative methods and strategies that enable tackling these fundamental open problems in the understanding of nature \cite{MCB2020review}. One alternative that is attracting increasing attention is the pathway of Tensor Network (TN) \cite{WhiteDMRG92,RomanReview2019,Simone2019BookReview}. TN is one of the mainstream paradigms for simulating quantum many-body lattice systems, both in and out of equilibrium, via a representation of the quantum state with tailored variational ansatz wave-functions. Originally introduced in the context of condensed matter physics, TN can solve quasi-exactly one dimensional strongly correlated quantum many-body problems for system sizes much larger than exact diagonalisation allows. They originated from the understanding that the density matrix renormalisation group (DMRG) technique could be recast in a variational formulation by means of Matrix Product States (MPS) \cite{OstlundRommer97,IgnacioFrankPBCMPS,UliAgeofMPS}. This stimulated the further development of such a framework in the last decade, extending the TN paradigm to encompass higher dimensionality \cite{PEPS2006,2DMERA,GersterFQH2017,CorbozIPEPSHeisembergSU4,LauchliIPEPS2018,MCBiPEPSgaugeZ3}, peculiar geometries \cite{HolographicTN,HolographicSpinTN}, directly addressing the thermodynamical limit \cite{IMPSComparison,IMPSTruncation,BinderDispersion}, as well as the continuum \cite{CMPS2010,CMPSVidal2017,ContinuousPEPSIgnacio19,ContinuousMERAFrank13}.

One of the most appealing features portrayed by TN is the possibility to encode and control global symmetries for the local degrees of freedom that characterise several condensed matter models \cite{TNSymmetrySingh,TNSymmetryWeichselbaum}. In fact, a general, robust and numerically efficient formulation of any such symmetries in the TN framework is known; it is commonly used in simulation to achieve an enhancement of the algorithm performance, as well as precise targeting of irreducible representation sectors \cite{TNSymmetrySingh2,TNAnthology,TNSymmetryWeichselbaum2}.

Lattice gauge symmetries differ from global ones, since they have quasi-local supports and are typically homogeneous, yielding a combined Lie algebra of generators which grows extensively with the system size \cite{WilsonConfinement,KogutSusskind75}. Nevertheless, several physical contexts have been found where TN are an exact description of the ground states of gauge-invariant Hamiltonians (e.g., 2D toric code that is an Ising gauge theory) \cite{ToricCode}. More recently, this framework has been successfully applied to LGT related problems. In fact TN represent microscopically the local Hilbert spaces and at the same time are tailored on a real-space wave-function representation, so they can be used to describe real-space locality and local symmetries altogether \cite{Tagliacozzo_Magnets, LGTN, GaugingGlobalLocal2015, ZoharBurrelloNJP}.

As previously mentioned, TN is naturally free from the sign problem \cite{TNvsSignProblem}. In fact, for 1-dimensional systems a number of successful studies have demonstrated the power of TN for HEP problems \cite{SugiharaPhi4DMRG,SugiharaZ2MPS,PsiPhi4} and lattice gauge theory calculations \cite{banuls2013mass,MCBSchwinger2014, Buyens2014, MCBNonAbelian2015, MCBChiralCondensate,Buyens_CutoffControl,Banuls2019}.
In particular, it has been shown that TN provide accurate determinations of mass spectra and that they can map out a broad temperature region. They can also treat chemical potentials and topological terms and they have been used to study realtime dynamics \cite{ThomasStringBreak,MCBNonAbelian2015,BuyensRealtime17}. TN also allows the study of entanglement properties and the entropy (leading in turn to the determination of central charges) in gauge theories, which brings new aspects of gauge theories into focus \cite{FerdinandSU2, YannickSU3}.  By quantitatively validating quantum simulators in out-of-equilibrium situations, even if only in lower dimensions, MPS and TN methods play a very important role towards establishing quantum simulators as reliable tools in quantum physics. Finally, recently the applications of TN to higher-dimensional LGT has start becoming within reach of the available algorithms and numerical resources~\cite{Corboz2014,Ran2017,He2018,Vlaar2021,TTN_QED2D, Magnifico_QED3D,Felser2021}, providing an important stimulus to further develop these techniques and explore their application to LGT.

In the following sections, we will review tensor network methods as a new tool for classical computation, fixing our attention and description on matrix product states and tree tensor networks. We will describe the Hamiltonian formulation of lattice gauge theories (albeit non-Hamiltonian approaches are possible \cite{meurice2020tensor}) as the set of models and theories to be treated with these tensor networks. Specifically, we will show the rishon formulation of LGT (a finite dimensional representation of the gauge group) which is specially well suited for the characterisation with TN. Then, we will present the ground state search with loop-free tensor networks for LGT simulation and we will show how can be used to study the confinement properties of a three dimensional compact QED. We will close the review with conclusions and perspectives.

\section{Tensor Network Methods: a flexible tool for classical computations}

From an information theory standpoint, Tensor Networks are complex, scalable data structures, whose information is stored within a collection of {\it tensors}: multi-indexed arrays $\mathcal{T}_{i_1, i_2, \ldots i_r}$ of real or complex floating-point numbers (in single-or more often double-precision). Tensor Networks represent a convenient way to store huge amount of data in a compressed, manageable, and approximate format. In fact, while the rank $r$ of a single tensor does {\bf not} scale with the size of the problem, tensors connect to each other via the {\it contraction operation}, generalising the matrix-matrix row by column multiplication operation, as
\begin{equation}
\mathcal{T}''_{i_1 \ldots i_r, j_1 \ldots j_s} = \sum_{k_1 \ldots k_p} \mathcal{T}'_{i_1 \ldots i_r, k_1 \ldots k_p}
 \mathcal{T}_{k_1 \ldots k_p,j_1 \ldots j_s}
\end{equation}
represented in the {\it network} as a link connecting two tensors $\mathcal{T}$ and $\mathcal{T}'$, which sit at the nodes of the network graph. Then, the size of the network itself scales with the size of the problem, and quickly the Tensor Network conveniently describes data structures which could not otherwise be stored in their exact, `fully contracted' form.

\begin{figure}
\centering
\begin{minipage}{.5\linewidth}
  \centering
  \includegraphics[width=0.9\linewidth]{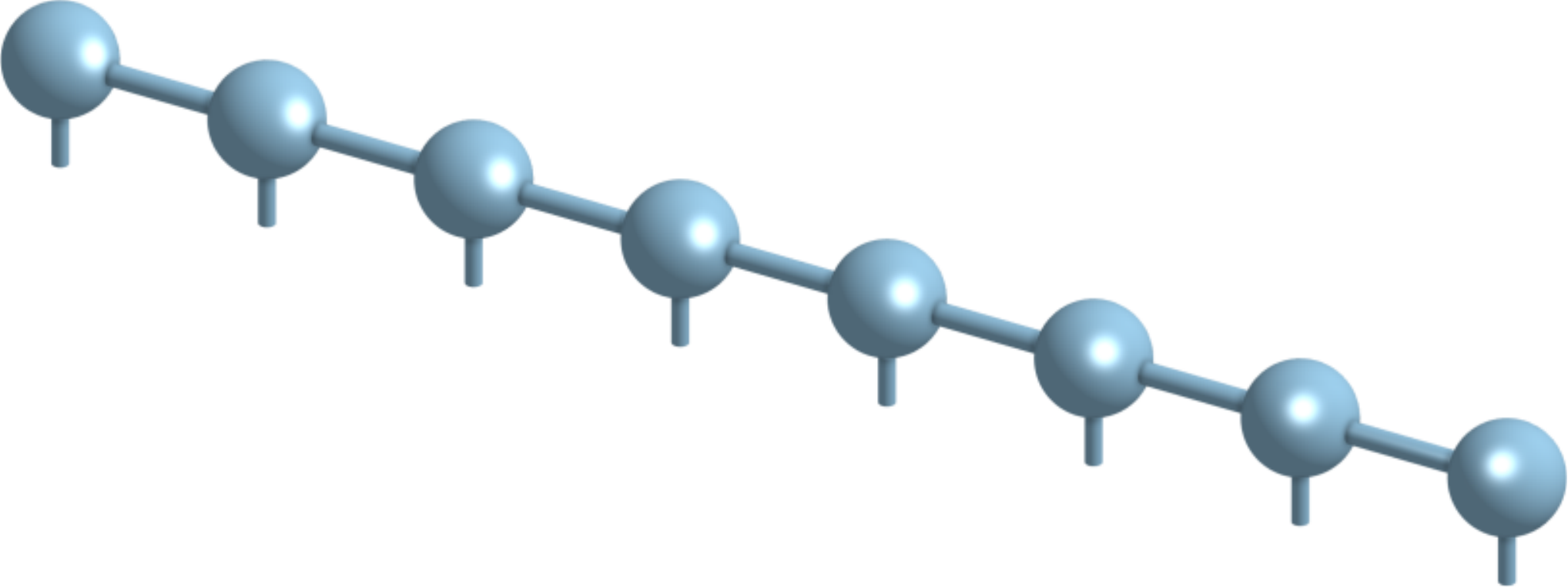}
  \\{Matrix Product States}
  \label{fig:Tagging}
\end{minipage}%
\begin{minipage}{.5\linewidth}
  \centering
  \includegraphics[width=0.9\linewidth]{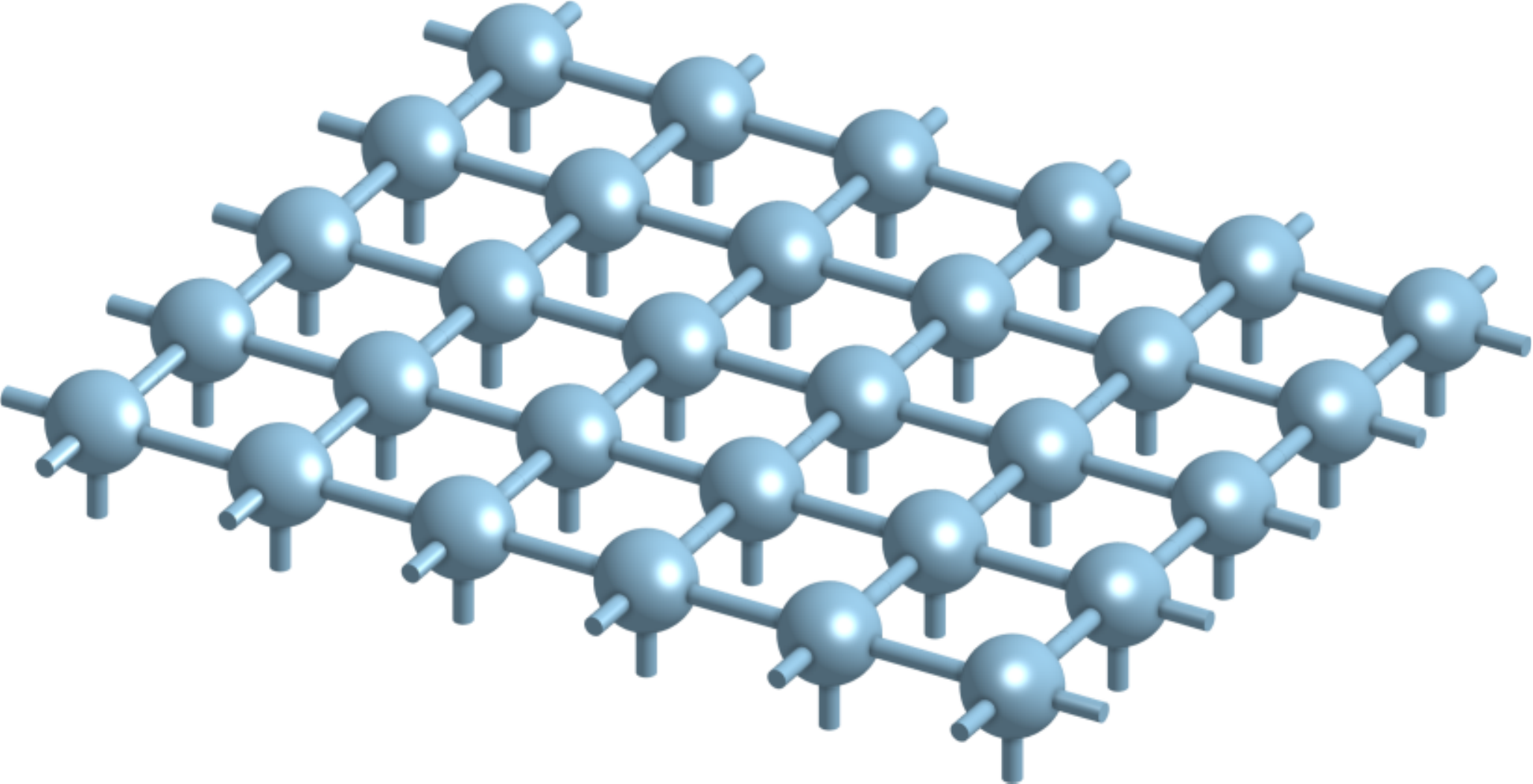}
  \\{Projected Entangled Pair States (2D)}
  \label{fig:probCorr}
\end{minipage}
\begin{minipage}{.5\linewidth}
  \centering
  \includegraphics[width=0.9\linewidth]{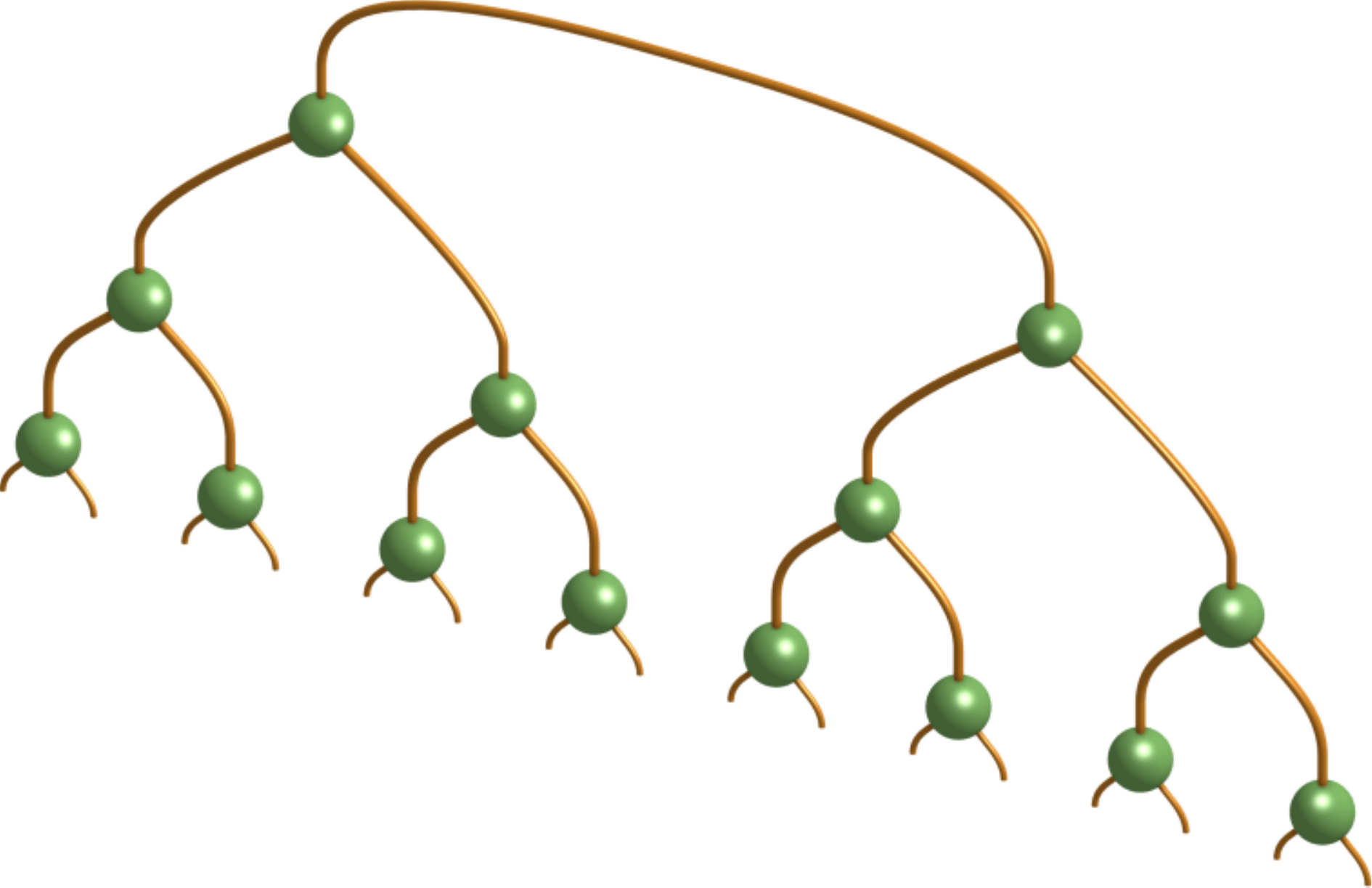}
  \\{Tree Tensor Network (1D)}
  \label{fig:probDNN}
\end{minipage}%
\begin{minipage}{.5\linewidth}
  \centering
  \includegraphics[width=0.6\linewidth]{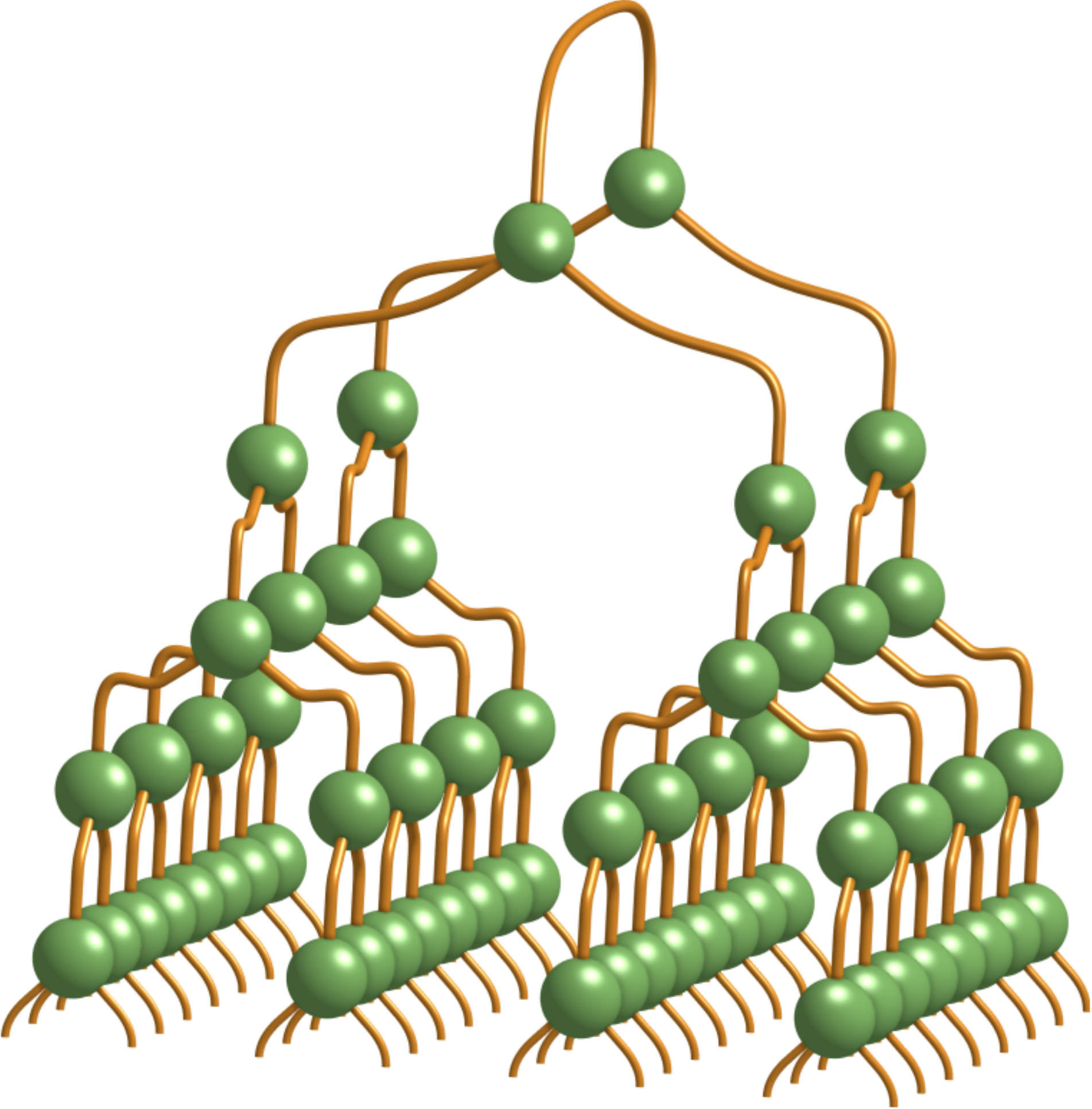}
  \\{Tree Tensor Networks (2D)}
  \end{minipage}
\caption{
Visual representations for common Tensor Network geometries, respectively for one-dimensional lattices (left panels) and for two-dimensional square lattices (right panels).
Bullets represent tensors, while each link represents a specific tensor-index, shared solely by the two tensors being connected.
Figure originally presented in Ref.~\cite{TTN_QED2D}, courtesy of the authors.}
\label{fig:probTTN}
\end{figure}

The standard application of TNs is the numerical simulations of Many-Body Quantum Systems (MBQS), where the variational many-body wave-function is encoded as a TN. Precisely, any state $|\Psi\rangle$ of a MBQS is defined by its amplitudes tensor
$\mathcal{T}{s_1 \ldots s_N}$, apart from a global phase $\mathcal{T} \to e^{i \varphi} \mathcal{T}$, via
\begin{equation}
 |\Psi\rangle = \sum_{s_1 \ldots s_N} \mathcal{T}{s_1 \ldots s_N} |s_1\rangle \otimes |s_2\rangle \otimes \cdots \otimes |s_N\rangle,
\end{equation}
where $|s_1\rangle$ is the logical basis (of dimension $d \geq 2$) of a single constituent of the system quantum, or {\it site}. As the size of $\mathcal{T}$ scales exponentially $\sim d^{N}$ with the system size $N$, for $N$ above few tens, it becomes mandatory to represent $\mathcal{T}$ in a compressed form, e.g.~a tensor network. Remarkably, the approximation necessarily introduced when compressing such exponentially large information into polynomially-scaling amount of data, is accurate when representing a large class of {\it realistic physical states}. For these many-body quantum states the entanglement -- that defines the maximal rate of possible compression without loss of information -- is relatively low and distributed according to the natural connectivity of the MBQS. These class of states, which satisfy the {\it area laws of entanglement} \cite{AreaLawReview} are naturally encoded in tensor networks and the error introduced by the TN approximation can be controlled. The degree of approximation introduced by a TN ansatz state is controlled by the {\it virtual bond dimension} $\chi$, i.e., the (largest) dimension $\chi \geq d$ among internal {\it virtual indices}, those indices that are contracted in order to obtain the amplitudes vector $\mathcal{T}{s_1 \ldots s_N}$. Specifically, for $\chi = d^{N}$ no approximation is introduced, and the description is exact. On the opposite limit, $\chi =1$ correspond to a mean-field description of the many-body state. In conclusion, TN methods are considered successful when a bond dimension $\chi$ scaling polynomially with $N$ is sufficient to achieve convergence in the physical quantities of interest. Simple prototypical TN are drawn in Fig.~\ref{fig:probTTN}.

From such perspective, developing Tensor Network algorithms consists in translating the well-understood protocols for $|\Psi\rangle$, ranging from equilibrium problems to real-time evolution, into protocols for the tensors in the TN that represents $|\Psi\rangle$. This framework has lead to the development of some of the most successful numerical methods for non-integrable lattice systems, including HEP problems: from DMRG \cite{WhiteDMRG92,UliAgeofMPS,ByrnesDMRGScwinger2002,SugiharaPhi4DMRG} to the Time-Evolved-Block decimation \cite{TEBD,TDMRG} and the Time-Dependent Variational Principle \cite{TDVP}, which are based on the Matrix Product State ansatz class \cite{OstlundRommer97,IgnacioFrankPBCMPS}, to the Tensor-Renormalisation-Group \cite{TRG}, which is a contraction method for infinite PEPS \cite{PEPS2006} and similarly enables the (approximate) calculation of lattice path integrals required in Lagrangian HEP problems \cite{YoshimuraTRGPathInt1,YoshimuraTRGPathInt2}.
Tensor Networks are also a physically meaningful ansatz in the context of Variational Monte Carlo methods \cite{TNMC07}, thus providing yet another pathway to study lattice gauge models in higher dimension \cite{ErezTNMC1,ErezTNMC2}.
Hereafter, we concentrate on loop-free TN, i.e.~TN displaying no cycles in their network architecture, namely MPS and Tree Tensor Networks (TTN).

\subsection{Matrix Product States}

The methods that we describe in what follows, share as a building block a set of local finite matrices that encode the information of the system. We show how to build these matrices, their properties and how they can be used to compute efficiently any correlator. We define the matrix product state showing two possible ways to built it. The first one is in the spirit of DMRG method of S. White \cite{WhiteDMRG92}, based on the Schmidt decomposition of a quantum state. The second one, the original point of view, comes from the fact that the matrices can be seen as a bond between the neighbour sites in a chain. The latter approach springs from the valence bond model introduced by I. Affleck, T. Kennedy, E.H. Lieb and H. Tasaki \cite{AKLT} and generalised by M. Fannes, B. Nachtergaele and R. F. Werner \cite{Fannes92}. The link between the two points of view has been unveiled by S. Ostlund and S. Rommer \cite{OstlundRommer97}. The results on the scaling of the entanglement faithfully represented by the ansatz lays at the heart of the success of this formalism in one dimensional system. Starting from these seminal works, it has been shown how to generalise them to nonlocal interactions, different boundary conditions and how to perform dynamical simulations of quantum systems with classical resources.

We first review some useful mathematical properties of composed quantum systems and show their implications for the MPS ansatz. The Schmidt decomposition of a quantum state between the first site and rest of the chain reads,
\begin{equation}
|\psi \rangle = \sum_{\alpha_{1} =1}^{D} \mu_{\alpha_{1}} |\alpha_{1}^{(1)} \rangle |\alpha_{1}^{(2 \cdots N)} \rangle 
\end{equation}
where $\{ |\alpha_{1}^{(1)} \rangle \}$ is an orthonormal basis of the first subsystem and $\{  |\alpha_{1}^{(2 \cdots N)} \rangle \}$ is an orthonormal set for the rest of the chain, the real numbers $\{ \mu_{\alpha_{1}} | 0 \le \mu_{\alpha_{1}} \le 1, \sum_{\alpha_{1} =1}^{D} |\mu_{\alpha_{1}}|^{2} =1 \}$ are the Schmidt coefficients. Finally, the integer $D$ is the Schmidt rank in the decomposition that, for easy writing, we fix to the maximum of any bipartition. If we express the state of the first site in some initial local basis, then the state is written,
\begin{equation}
|\psi \rangle = \sum_{\alpha_{1} =1}^{D} \sum_{s_{1}=1}^{d}  \mu_{\alpha_{1}} |s_{1}\rangle \langle s_{1} |\alpha_{1}^{(1)} \rangle |\alpha_{1}^{(2 \cdots N)} \rangle =  \sum_{\alpha_{1} =1}^{D} \sum_{s_{1}=1}^{d}  |s_{1}\rangle A^{s_{1}}_{\alpha_{1}}  |\alpha_{1}^{(2 \cdots N)} \rangle
\end{equation}
where in this case, $d$ is the dimension of the local space and $A^{s_{1}}_{\alpha_{1}} = \mu_{\alpha_{1}} \langle s_{1} |\alpha_{1}^{(1)} \rangle $ is a tensor describing the first site.

Repeating this manipulation for the second site, we find
\begin{equation}
|\psi \rangle  = \sum_{\{\alpha\}=1}^D  \sum_{\{s\}=1}^{d} |s_1\rangle  A_{\alpha_1}^{s_1} ~ |s_2\rangle   A_{\alpha_1\alpha_2}^{s_2} |\alpha_2^{(3...N)}\rangle ,
\end{equation}
where the sums $\{\alpha\}$ and $\{s\}$ are made over the different configurations of $\alpha_1$, $\alpha_2$, $s_1$ and $s_2$ and the matrix  $A_{\alpha_1\alpha_2}^{s_2} =  \mu_{\alpha_2} \langle  s_2|\alpha_1\alpha_2^{(2)}\rangle $. Then for the whole chain, the state is written as follows,
\begin{equation}
|\psi\rangle =  \sum_{\{\alpha\}=1}^D  \sum_{\{s\}=1}^{d} |s_1\rangle  A_{\alpha_1}^{s_1} ~ |s_2\rangle   A_{\alpha_1\alpha_2}^{s_2} \cdots |s_{N-1}\rangle  A_{\alpha_{N-2}\alpha_{N-1}}^{s_{N-1}} |s_N\rangle  A_{\alpha_{N-1}}^{s_N}.
\end{equation}

Looking at the bulk of an infinite chain, the basis in the bipartition for the right subsystem is written as,
\begin{equation}
|\alpha_{L-1}\rangle = \sum_{\alpha_L =1}^D \sum_{s_L=1}^d |s_L\rangle  A_{\alpha_{L-1} \alpha_L}^{s_L} |\alpha_L\rangle , 
\end{equation}
where $|\alpha_L\rangle = \sum_{\alpha_{L+1} =1}^D \sum_{s_{L+1}=1}^d |s_{L+1}\rangle  A_{\alpha_L \alpha_{L+1}}^{s_{L+1}} |\alpha_{L+1}\rangle $ and if $\{|\alpha_{L-1}\rangle \}$ is an orthonormal basis then 
\begin{equation}
\sum_{s_L=1}^d \sum_{\alpha_L=1}^D \left(A_{\alpha'_{L-1} \alpha_L}^{s_L}\right)^* A_{\alpha_{L-1} \alpha_L}^{s_L} =\sum_{s_L=1}^d  \left(A^{s_L}\right)^{\dagger} A^{s_L} =\mathbb{I},
\end{equation}
which defines the condition in the $A^s$ matrices to be a positive, trace preserving map. Within this approach, the matrices $A^{s_L}$ represent the change in the description of a system when it is considered one site more, they act as a transfer matrix. Every time we insert one site more in the subsystem, a new matrix $A^{s_L}$ link it with the rest of the chain and, in that way, a succession of matrices $\{A^{s_L}\}$ is produced. If this succession has a limit, then we could represent the final state of the chain by this set of matrices. 

This description is computationally efficient if the Schmidt rank $D$ is bounded. This situation appears usually in one dimensional system with local interactions. When we perform the bipartition of the system between L sites and the rest of the chain, the Schmidt rank measures the number of degree of freedom needed to describe the subsystem and is always bounded by the entropy of entanglement, i.e., $D \gtrsim 2^{S_L}$. Using the analysis of the scaling of entanglement in one dimensional models, we know that for finite systems of length $N$ the maximum of the entropy goes as $S_L\sim \log_2{N}$ and for infinite gapped systems, it goes like $S_L\sim \log_2{\frac{1}{m}}$. So, in any of these case the Schmidt rank is bounded by a constant. Even at the critical points, we know that the entropy grows with the size of the subsystem $L$ as $S_L\sim \log_2{L}$, then, although the Schmidt rank is not saturated in this case, its growth is polynomial with the size of the subsystem, $D\sim L$.

In a translationally invariant state, the whole set of matrices can be represented by just one matrix independently of its position, i.e.,  $\{A^{s_L}\} \to A^s$. Then, a translationally invariant state can be described by the product of this matrix as follows,
\begin{equation}
|\psi\rangle =\sum_{\{s\}=1}^d  \text{Tr} \left( B \cdot A^{s_1} \cdots A^{s_N} \right) |s_1, \cdots, s_N  \rangle 
\end{equation}
where the matrix $B$ implements the boundary conditions of the system. In particular, if the state has periodic boundary condition this matrix is the identity, $B=\mathbb{I}$.

The second point of view, for the matrix product state, comes from the identification of the lower indexes of the matrix $A_{\alpha \beta}^{s}$ with two ancillary subsystems used to implement the state of the physical system $s$. This one was the original method due to Affleck, Kennedy, Lieb and Tasaki (AKLT) to built the exact ground state of antiferromagnetic system of a spin one chain using the symmetric subspace of two spin $1/2$ ancillae.
The matrix appears as a projector from the Hilbert space $\mathcal{H}^{(a)}$ of the ancillae to the real one $\mathcal{H}^{(phys)}$, i.e.,
\begin{equation}
\begin{split}
A:~& \mathcal{C}^D \otimes \mathcal{C}^D \to \mathcal{C}^d,\\
&|\alpha,\beta\rangle \to A|\alpha,\beta\rangle= \sum_{s=1}^d  A_{\alpha \beta}^{s} |s\rangle.
\end{split}
\end{equation}
The correlations in the chain are implemented by maximally entangled states in the ancillae subsystem of neighbour site, i.e. $\sum_{\alpha=1}^D |\alpha\rangle |\alpha\rangle $, and the final description of the state is,
\begin{equation}
|\psi \rangle  = A \sum_{\alpha=1}^D |\alpha\rangle  |\alpha\rangle \sum_{\beta=1}^D |\beta\rangle |\beta\rangle  = \sum_{\alpha,\beta=1}^D \sum_{s=1}^d |\alpha\rangle   A_{\alpha \beta}^{s} |s\rangle |\beta\rangle .
\end{equation}

Once the state of the system is built, it is important to be able to compute any correlator from it. In the matrix product state formalism, for translationally invariant systems, this task is implemented defining a transfer matrix. For instance, to get the normalisation in a translational invariant state,
\begin{equation}
\langle \psi | \psi \rangle  =  \sum_{\{s\}=1}^d   \sum_{\{\tilde{s}\}=1}^d  \langle  s_1 .. s_N| \text{Tr} \left( A^{s_1} .. A^{s_N} \right)^*  \text{Tr} \left( A^{\tilde{s}_1} .. A^{\tilde{s}_N} \right) |\tilde{s}_1.. \tilde{s}_N  \rangle ,
\end{equation}
defining the $D^2 \times D^2$ transfer matrix $E=\sum_{s=1}^d  (A^{s})^* \otimes A^{s}$, the normalisation is just $\langle  \Psi | \Psi \rangle  = \text{Tr}  \left(E^N \right)$. If the state is normalised, then, the spectrum of the matrices $E$ is such that $\{\lambda_{\mu} | ~|\lambda_{\mu}|\le1,  ~1 \le \mu \le D^2 \}$. If the state fulfils the clustering principle\footnote{This principle requires that $\lim_{x \to \infty}{
\langle  \psi |\mathcal{O}(x) \mathcal{O}(0) |\psi \rangle }= \langle  \psi |\mathcal{O}(x) |\psi\rangle  \langle  \psi| \mathcal{O}(0) |\psi \rangle $}, then, the maximum eigenvalue $\lambda_1=1$ is unique.

In this representation the two point function of any two operators at the site $(i)$ and $(j)$ is written as
\begin{equation}
\langle  O^{(i)} O^{(j)} \rangle  = \text{Tr}  \left(E^{N-j+i-1} \tilde{O}^{(i)} E^{j-i-1} \tilde{O}^{(j)} \right)
\end{equation}
where $\tilde{O}=\sum_{\{s\}=1}^d \sum_{\{\tilde{s}\}=1}^d  (A^{s})^* \otimes A^{\tilde{s}} \langle s| O | \tilde{s}\rangle$. If we consider an infinite chain and taking the diagonal form of $E=|1\rangle \langle 1|+\sum_{\mu=2}^{D^2}\lambda_{\mu} |\mu\rangle \langle \mu |$, then,
\begin{equation}
\begin{split}
\langle  O^{(i)} O^{(j)} \rangle & - \langle  O^{(i)} \rangle  \langle  O^{(j)} \rangle   = \langle  1| \tilde{O}^{(i)} \left(\sum_{\mu=2}^{D^2} (\lambda_{\mu})^{j-i-1} |\mu\rangle  \langle  \mu | \right) \tilde{O}^{(j)} |1\rangle  \\
&= \sum_{\mu=2}^{D^2} \langle  1| \tilde{O}^{(i)} |\mu\rangle  \langle  \mu |\tilde{O}^{(j)} |1\rangle  \left(\frac{\lambda_{\mu}}{|\lambda_{\mu}|}\right)^{j-i-1} e^{-\frac{j-i-1}{\xi_{\mu}}},
\end{split}
\end{equation}
where $\xi_{\mu}=\frac{-1}{\log{|\lambda_{\mu}|}}$ defines the correlation length. From this expression, we see that the long range behaviour of the system is implement in the eigenvalues of the transfer matrix $E$, that is invariant under the redefinition of the local basis. 

\subsection{Tree Tensor Networks}

Within a zoo of tensor network architectures, the TTN holds a special mention in versatility. TTN algorithms can be implemented on lattice of {\it any} spatial dimensionality (from 1D to 3+D) with the same computational scaling cost, of order $O(\chi^4)$. The TTN has tensors (or graph nodes) organised in layers. Every tensor merges together two neighbouring `effective sites' into a single one that then proceeds to the next layer. Ultimately, the TTN becomes completely connected once we include $N-2$ Tensors, with three indexes each, arranged into $\log_2 N$ layers, regardless of the spatial dimension. The computational simplicity of such a TN ansatz carries a core drawback: the incapability of capturing the correct scaling of microscopic entanglement for fixed $\chi$. Nevertheless, in a context where efficient algorithms for sophisticated TNs (appropriate for 3+1D) are still under development, the TTN ansatz is still a great accuracy improvement over (cluster) mean-field approaches, and is especially useful for models affected by the sign problem, where Monte Carlo strategies perform poorly, as we will see in the following sections.

As a loop-free (or loop-less) tensor network, TTNs benefit from an extremely robust groundwork of algorithms, developed and refined over two decades, capable of performing the two standard tasks of (closed-system) quantum mechanics: {\it (i)} ground state search, and {\it (ii)} real-time evolution, corresponding respectively to the time-independent and time-dependent formulations of the Schr{\"o}dinger equation, for a many-body lattice Hamiltonian $H$. Modern TTN numerical suites employ a generalisation of DMRG for loop-less tensor networks in order to tackle the ground-state search \cite{GersterUnconstrained,TNAnthology}, while real-time dynamics is typically carried out by Time-Dependent Variational Principle (TDVP), originally designed for MPS \cite{TDVP,TDVP2}, but allowing a natural generalisation for loop-less TN \cite{TDVPTTNLukas,TDVPChin}.

\section{Lattice Gauge Theories in Hamiltonian formulation}

Tensor Networks methods have been extensively developed to tackle interacting condensed matter problems and strongly correlated many-body quantum problem. Their technology is centered around casting the problem of dynamics in Hamiltonian formulations (or its open-system generalisations, such as the master equation for Markovian dynamics). In this context, the most natural way of addressing high-energy problems via TN is to cast them in a lattice Hamiltonian formulation, where the time dimension (or temperature dimension) remains a continuous variable, while the space is discretised in a (square or cubic) lattice. While this approach inevitably breaks Lorentz invariance (continuous scale transformations are not lattice isomorphisms) it has the benefit of removing any possible infrared or ultraviolet divergencies that may arise in the continuous theory, especially for finite system sizes. As the lattice spacing $a$ is a simulation parameter that can be fully controlled, the overall behaviour of the continuum theory emerges in the limit $a \to 0$, where Lorentz symmetry is expected to be restored.

In Hamiltonian Lattice Gauge theories \cite{KogutSusskind75}, gauge fields are defined on the bonds of the lattice, while matter fields are defined on the vertices of the lattice. The matter fields are meant as relativistic particles, and they mutually interact {\it only} via the gauge fields. Therefore, in the limit of zero gauge coupling ($g_c = 0$), the matter field Hamiltonian describes a free lattice Dirac dynamics. While this is not the unique way to do so (see for example Wilson fermions \cite{WilsonConfinement,GinspargWilson82}), Susskind lattice fermions are often adopted in this context \cite{SusskindLatticeFermions}, where in the limit $a \to 0$ the quadratic Hamiltonian
\begin{equation} \label{eq:susskind}
 \begin{aligned}
 H_{\text{Susskind}} &= m c^2 \sum_{\jvec} (-1)^{j_x + j_y + j_z} \psi^{\dagger}_\jvec \psi_\jvec + \\
  & \quad + \frac{c \hbar}{2 a} \sum_{\jvec} \left(
	-i \psi^{\dagger}_\jvec \psi_{\jvec + \mu_x}
	-(-1)^{j_x + j_y} \psi^{\dagger}_\jvec \psi_{\jvec + \mu_y}
	-i(-1)^{j_x + j_y} \psi^{\dagger}_\jvec \psi_{\jvec + \mu_z}
	+ \text{H.c.} \right)
 ,
 \end{aligned}
\end{equation}
converges to free Dirac dynamics $H_{\text{Dirac}} = c \gamma^{0} \vec{\gamma} \cdot \vec{p} + m c^2 \gamma^{0}$ (albeit with potentially degenerate solutions a.k.a.~fermion doubling, depending on the spatial dimension). The Fermi field $\psi_\jvec$, satisfying $\{\psi_\jvec, \psi_{\jvec'}\} = 0$ and $\{\psi_\jvec, \psi^{\dagger}_{\jvec'}\} = \delta_{\jvec, \jvec'}$ anti-commutation relations, has a basis in the lattice, each site within the supercell corresponding to one component of the Dirac spinor (e.g. basis of 4-sites in a supercell for 3+1D, that matches the Dirac 4-spinor).

When the gauge coupling is not zero the hopping terms are modified to include gauge field operators, in such a way that the resulting term is gauge-invariant at both sites. Two prominent examples of such a procedure are Quantum Electrodynamics (QED) and Quantum Chromodynamics (QCD), each deserving its own characterisation.

\subsection{QED}

The local symmetry group of QED is U$(1)$ (parametrised by a single angle $\varphi_{\jvec}$ per site), and the gauge degree of freedom (DoF) takes into account the presence or absence of a (charged) particle. Thus, matter fields gauge-transform by gaining a site-dependent phase $\psi_j \to e^{i \varphi_{\jvec}} \psi_j$. To restore gauge symmetry invariance of the hopping terms in Eq.~\eqref{eq:susskind} we introduce additional local degrees of freedom, or `lattice gauge fields', living on the bonds of the lattice. Hopping terms are then modified according to
\begin{equation}
 \psi^{\dagger}_\jvec \psi_{\jvec + \mu} \longrightarrow \psi^{\dagger}_\jvec U_{\jvec,\jvec+\mu} \psi_{\jvec + \mu}
\end{equation}
along each direction $\mu = \{\mu_x, \mu_y, \mu_z\}$, where the operator $U_{\jvec,\jvec+\mu}$, in order to protect gauge symmetry both at site $\jvec$ and $\jvec+\mu$, transforms as $U_{\jvec,\jvec+\mu} \to e^{i( \varphi_{\jvec} - \varphi_{\jvec +\mu})} U_{\jvec,\jvec+\mu}$ . On top of this, we add the Hamiltonian of pure electrodynamics, allowing photons to carry energy and propagate, which is expected to converge to $\frac{1}{4} F_{\mu \nu} F^{\mu \nu}$ in the continuum limit. A common lattice description is the one developed by Kogut and Susskind \cite{KogutSusskind75},
\begin{equation} \label{eq:pureQED}
 \begin{aligned}
  H_{\text{pure,QED}} &= \frac{c \hbar}{2a} g_{\text{em}}^{2} \sum_{\jvec,\mu} L^2_{\jvec,\jvec+\mu} \\
	&+ \frac{4 c \hbar}{a g_{\text{em}}^{2}} \sum_{\jvec,\mu \neq \mu'} \left( U_{\jvec,\jvec+\mu} U_{\jvec+\mu,\jvec+\mu+\mu'} U^{\dagger}_{\jvec+\mu',\jvec+\mu+\mu'} U^{\dagger}_{\jvec,\jvec+\mu'}
	+ \text{H.c.} \right),
 \end{aligned}
\end{equation}
where $g_{\text{em}}$ is the universal, dimensionless electromagnetic coupling. Such pure Hamiltonian has the advantages of being explicitly gauge-invariant and short-ranged, despite the drawback of displaying a four-bonds interaction, a.k.a. the {\it plaquette term}. The bond-operators $L_{\jvec,\jvec+\mu} = L^{\dagger}_{\jvec,\jvec+\mu}$ and $U_{\jvec,\jvec+\mu}$ define the gauge DoF space via their Lie algebra, which reads $[L_{\jvec,\jvec+\mu},U_{\jvec',\jvec'+\mu'}] = \delta_{\jvec,\jvec'} \delta_{\mu,\mu'} U_{\jvec,\jvec+\mu}$ and $[U^{\dagger}_{\jvec,\jvec+\mu},U_{\jvec',\jvec'+\mu'}] = 0$. Basically $L_{\jvec,\jvec+\mu}$ counts the number of electric field quanta (with direction encoded in the sign) on the bond, while $U^{\dagger}_{\jvec,\jvec+\mu}$ shifts this number by one quantum. Notably, exact representations of such an algebra are infinite-dimensional, as rightfully expected from the gauge-field being bosonic in nature. We therefore have to introduce an approximation (an energy-density cutoff) to effectively work with the finite dimensional-local spaces required by Tensor Network methods. Proper limits to the infinite dimensional representations has been shown to converge in different cases~\cite{Banuls2019}.  

\subsection{QCD}

In this scenario, the local symmetry group is non Abelian, corresponding to SU$(3)$. Matter fields $\psi_{\jvec c}$ carry an internal degree of freedom, the {\it color} $c \in \{\str, \stg, \stb \}$, that unlike spin does not couple to their momentum in the free Dirac equation. The color DoF is manipulated by gauge-transformations, which read $\psi_{\jvec c} \to \sum_{c'} \left[ \exp( {i \sum_{\tau}^8 \varphi_{\jvec \tau} \Lambda_{\tau}} ) \right]_{c c'} \psi_{\jvec c'}$, where $\Lambda_{\tau}$, with $\tau \in \{ 1 .. 8 \}$, are the eight $3 \times 3$ Gell-Mann matrices. Here, a gauge transformation is uniquely defined by an octuplet of angles $\varphi_{\jvec \tau}$ for each site $\jvec$. Analogously to the QED case, in order to restore gauge invariance, we introduce operators, acting on the gauge degrees of freedom on the links, that modify the hopping terms, according to
\begin{equation}
 \psi^{\dagger}_\jvec \psi_{\jvec + \mu} \longrightarrow \psi^{\dagger}_{\jvec c} U_{\jvec,\jvec+\mu;c c'} \psi_{\jvec + \mu,c'},
\end{equation}
where the shift (or comparator) operator $U_{\jvec,\jvec+\mu;c c'}$ transforms as
\begin{equation}
 U_{\jvec,\jvec+\mu;c c'} \to \sum_{b, b' \in \{\str \stg \stb \}}
\left[ \exp( {+i \sum_{\tau}^8 \varphi_{\jvec \tau} \Lambda_{\tau}} ) \right]_{c b}
\left[ \exp( {-i \sum_{\tau}^8 \varphi_{\jvec+\mu,\tau} \Lambda_{\tau}} ) \right]_{c' b'} U_{\jvec,\jvec+\mu;b b'}
\end{equation}
to protect the gauge-symmetry. 

One may interpret the bond space as composed of two sub-sites, a.k.a. 'rishon' modes: The gauge transformation effectively SU(3)-rotates the left rishon with angles $\varphi_{\jvec \tau}$, and simultaneously SU(3)-rotates the right rishon with angles $\varphi_{\jvec+\mu, \tau}$. Thus effectively, each rishon carries a representation of the SU(3) group.

Analogously to the QED case, we need to provide a pure chromodynamics Hamiltonian, on top of the covariant matter dynamics, to recover the correct SU$(3)$ Yang-Mills dynamics in the continuum limit. Unsurprisingly, it can be written similarly to Eq.~\eqref{eq:pureQED}, precisely \cite{KogutSusskind75,Kogut_RMP83}
\begin{multline} \label{eq:pureQCD}
  H_{\text{pure,QCD}} = \frac{c \hbar}{2a} g_{\text{strong}}^{2} \sum_{\jvec,\mu} C^2_{\jvec,\jvec+\mu} \\
	+ \frac{4 c \hbar}{a g_{\text{strong}}^{2}} \sum_{\jvec,\mu \neq \mu'} \sum_{c_i \in \{\str \stg \stb\}}
	\left( U_{\jvec,\jvec+\mu; c_1 c_2} U_{\jvec+\mu,\jvec+\mu+\mu'; c_2 c_3} U^{\dagger}_{\jvec+\mu',\jvec+\mu+\mu'; c_4 c_3} U^{\dagger}_{\jvec,\jvec+\mu'; c_1 c_4}
	+ \text{H.c.} \right),
\end{multline}
with the strong force coupling $g_{\text{strong}}$. The 'electric' part of the Hamiltonian now displays the operator $C^2$: This is the quadratic Casimir operator of the SU(3) algebra applied to the rishon space (either the left rishon mode or the right rishon mode: their eigenvalues always match on physical states \cite{BurrelloZoharPRD}). Since SU(3) is a compact Lie group, the complete bond space is tied to the regular representation of SU(3), and thus intrinsically infinite-dimensional \cite{BurrelloZoharPRD}. Therefore it is mandatory to introduce approximations, or cutoffs, to manage numerical simulations of lattice QCD or other non-Abelian lattice gauge models \cite{Kogut_RMP79}

\section{Rishons and dressed logical bases}

At this stage, the only missing ingredient to enable numerical simulation of the aforementioned lattice gauge Hamiltonians via Tensor Network methods is the selection of a physically relevant finite-dimensional subspace of the lattice gauge field (bond space), to be included in the simulation. We stress that in one-spatial dimension, under open-boundary-conditions, this problem can be circumvented by `accumulating' the matter gauge-charges from one edge: This fact is notorious for the gauge-Abelian Schwinger model \cite{Martinez2016QSim}, but it has been recently shown to allow non-Abelian generalisations \cite{MCB_AccumulateSU2}. From two spatial dimension, this strategy cannot be exploited, and physically relevant cutoffs based on energy-density has to be introduced to access low-temperature lattice-gauge physics with TN, as routinely done when studying lattice models of interacting soft-core bosons \cite{GersterDisorderedBHM}.

A simple, yet meaningful choice of truncation is based on the `electrical' part of the pure lattice-gauge Hamiltonian, i.e.~the first row in Eqs.~\eqref{eq:pureQED} and \eqref{eq:pureQCD}. For each Rishon mode, only a selected number of irreducible representations (irreps) of the gauge group are fully included: those with quadratic Casimir value below a given cutoff, i.e.~states with weak electric gauge field. Other irreps are fully excluded. The nice aspect of such an approximation is that the cutoff itself can be tuned, thus numerical convergence in the cutoff can be systematically tested, which is especially useful when approaching the continuum limit \cite{Buyens_CutoffControl}.

Once the gauge field space has been truncated, since the matter field is intrinsically finite-dimensional due to quarks and leptons being fermions, TN simulations can be carried out without further obstacles. At this stage, it may be helpful to encode symmetries, both gauge and global, into the TN variational ansatz state, to ensure that the simulation does not explore unphysical states, while at the same time speeding-up the computation.

Once again, there is no single approach to tackle this task. Refs.~\cite{Tagliacozzo_Magnets,LGTN} offer a general, practical and efficient strategy, that displays several advantages. The basic idea is to group each matter field site $\jvec$ together with all its neighbouring rishon modes $(\jvec,\pm \mu)$, where the gauge field bond $(\jvec,\jvec+\mu)$ splits into rishon modes $(\jvec,\mu)$ and $(\jvec+\mu,-\mu)$, a {\it dressed site}. Interestingly, Gauss' Law, enforcing that physical state protect gauge symmetry, becomes a local selection rule on each single dressed site. It basically requires the dressed site, as a whole, to be always in the trivial irrep of the gauge group. This requirement effectively translates into the familiar Maxwell Equation $\vec{\nabla} \cdot \vec{E} - \rho = 0$ for QED (dressed sites must have effective charge zero), while in QCD it ensures that lonely quarks have always a gluon string attached to them, fostering confinement. As a local (potentially non-Abelian) 1-site selection rule, the trivial irrep requirement is effectively implemented by defining a local dressed logical basis (canonical basis for the dressed site) which includes only gauge-invariant states, effectively reducing the dimension of the dressed site space. We must be careful, however, that at any point in the simulation two rishons belonging to the same bond are in mutually adjoint irreps: This second requirement (Link Law) basically translates into a nearest-neighbour (2-dressed site) selection rule, which is {\bf always Abelian} even when the gauge-group is non-Abelian.

In conclusion, this strategy allows to `hide' all the non-Abelian content of the gauge theory (and the calculus of Clebsh-Gordan coefficients) in the definition of the dressed logical basis and the effective Hamiltonian operators over that basis. The Tensor Network algorithm handles the Abelian nearest-neighbour Link Laws, as well as the Abelian global symmetries (global charge conservation), both of which can be easily accounted for with the numerical TN suites available nowadays.

\subsection{QED dressed bases and defermionization}

An additional advantage of this approach emerges in the QED case. Due to the fact that the fusion rules of the gauge group ($U(1)$), equivalent to $\mathbb{Z}$, contain $\mathbb{Z}_2$ as a subgroup, it is possible to make rishon modes fermionic. In turn, this will produce an algebra of effective operators, over dressed logical bases, which are {\it genuinely local}, meaning that they commute at different sites, thus {\it defermionizing} the theory.

This technique, explored in Refs.~\cite{TTN_QED2D,ErezB}, will be briefly reviewed below for a special case. Precisely, we consider the scenario where we chose the electric energy cutoff to include three states of gauge field at each bond, namely $|1,-1\rangle$, $|0,0\rangle$ and $|-1,1\rangle$. In this format, we can write the bond-space operator algebra as
\begin{equation}
  U = |-1,1 \rangle \langle 0,0| + |0,0 \rangle \langle 1,-1|
	\qquad \mbox{and} \qquad
	L = |-1,1 \rangle \langle -1,1| - |1,-1 \rangle \langle 1,-1|
\end{equation}
which indeed satisfies the relation $[U,L]=U$. We can now decompose $U_{\jvec,\jvec+\mu}$ in a pair of fermionic rishons, specifically as $\xi_{\jvec,\mu} \xi^{\dagger}_{\jvec+\mu,-\mu}$. These rishons are exotic fermions, since they mutually anti-commute, i.e.~$\{ \xi_{\jvec,\mu}, \xi^{\dagger}_{\jvec',\mu'} \} = \{ \xi_{\jvec,\mu}, \xi_{\jvec',\mu'} \} = 0$ for $\jvec \neq \jvec'$ or $\mu \neq \mu'$, yet they are 3-hardcore, i.e.~$\xi_{\jvec,\mu}^3 = 0 \neq \xi_{\jvec,\mu}^2$, in contrast to standard (Dirac) fermions which are 2-hardcore.

In fact, it is possible to build such a rishon from two Dirac-fermion subspecies $a$ and $b$ at each site, with standard rules $\{ a_{\jvec,\mu}, a^{\dagger}_{\jvec',\mu'} \} = \{ b_{\jvec,\mu}, b^{\dagger}_{\jvec',\mu'} \} = \delta_{\jvec,\jvec'} \delta_{\mu,\mu'}$ and all the other anti-commutators zero. $\xi$ can then be written as
\begin{equation}
  \xi_{\jvec,\mu} = a^{\dagger}_{\jvec,\mu} a_{\jvec,\mu} b_{\jvec,\mu} + (1 - b^{\dagger}_{\jvec,\mu} b_{\jvec,\mu} ) a_{\jvec,\mu}.
 \end{equation}
The electric field operator $L$ can similarly be expressed as a function of these field operators, specifically
\begin{equation}
  2 L_{\jvec,\jvec+\mu} = a^{\dagger}_{\jvec,\mu} a_{\jvec,\mu} 
  + b^{\dagger}_{\jvec,\mu} b_{\jvec,\mu} 
  - a^{\dagger}_{\jvec+\mu,-\mu} a_{\jvec+\mu,-\mu}
  - b^{\dagger}_{\jvec+\mu,-\mu} b_{\jvec+\mu,-\mu}.
 \end{equation}
As we perform these substitutions into the lattice QED Hamiltonian, we observe that the total parity of fermions (rishons and matter) at each site is preserved by each and single term, i.e., it is a symmetry of the problem. In other words, each operator, acting on site $\jvec$, appearing in the Hamiltonian, effectively commutes with any other term in the Hamiltonian acting on site $\jvec'$: the mapping has {\it defermionized} the interacting many-body problem, as it can be equivalently expressed as a (large, interacting) spin-problem with the same interaction range (in this case, nearest-neighbour in the dressed sites).

While we presented explicit defermionization of three-gauge-states QED, we stress that a similar procedure is available for any arbitrary cutoff of the gauge field, thus it can be carried out at every energy scale without loss of generality. Moreover, similar strategies can be carried out whenever the fusion rules of the gauge group contain $\mathbb{Z}_2$ as a subgroup \cite{ErezA,ErezB}.

\section{Ground State Search with Loop-Free Tensor Networks for LGT simulation}

For the sake of compactness, we describe a practical architecture of loop free tensor networks, the binary tree tensor networks (bTTN), for general discussion see for instance Ref.~\cite{TNAnthology}. This network is built on top of a one-dimensional lattice of physical links. The bTTN is composed entirely of tensors with three links. The evaluation of local and two-point observables is performed as followed: The expectation value of a local observable $O^{[s]}$, acting on some site $s$ of a bTTN state $|\psi \rangle$, can be calculated by contracting just three tensors (independent of the system size $N$),
\begin{equation}
\langle \psi | O^{[s]} | \psi \rangle = \sum_{\alpha_{s} , \beta_{s} , \alpha_{(s+1)} , \alpha_{\zeta}  } \mathcal{T}^{[q]}_{\alpha_{s} , \alpha_{(s+1)} ,\alpha_{\zeta} } O^{[s]}_{\alpha_{s} , \beta_{s}}  \mathcal{T}^{[q]*}_{\beta_{s} , \alpha_{(s+1)} ,\alpha_{\zeta} }.
\end{equation}
and the two-point correlation observables
\begin{equation}
\langle \psi | O^{[s]} O'^{[s']} | \psi \rangle = \sum_{\alpha_{s} , \beta_{s} , \alpha_{(s+1)} , \alpha_{\zeta} , \beta_{\zeta}  } \mathcal{T}^{[q]}_{\alpha_{s} , \alpha_{(s+1)} ,\alpha_{\zeta} } O^{[s]}_{\alpha_{s} , \beta_{s}}  \tilde{O}'_{\alpha_{\zeta} , \beta_{\zeta}  } \mathcal{T}^{[q]*}_{\beta_{s} , \alpha_{(s+1)} ,\beta_{\zeta} }.
\end{equation}
where $\tilde{O}'$ is the result of the observable $O'^{[s']}$ obtained from contractions involving all tensors along the path between sites $s$ and $s'$ that can be evaluated with a very small number of contractions $\mathcal{O}(\log^{2} N)$, owing to the enhanced reachability of two arbitrary sites $s$ and $s'$ due to the hierarchical tree structure. This feature makes the calculation of observables a computational cheap task in a bTTN.

From the previous expectation values, the energy expectation value can be extracted and in particular, if a tensor in the whole tree structure is left free to optimise the energy expectation value, this can be recasted to an eigenvalue problem of the form
\begin{equation}
E= \langle  \mathcal{T}^{[q]} | \tilde{H}' |  \mathcal{T}^{[q]} \rangle
\end{equation}
with the constraint $\langle  \mathcal{T}^{[q]} |  \mathcal{T}^{[q]} \rangle =1$, where $\tilde{H}'$, in the same way as $\tilde{O}'$, is the result of the physical Hamiltonian with the contractions of all tensors but $ \mathcal{T}^{[q]} $. Now, the optimisation problem consists in determining the variational tensor $ \mathcal{T}^{[q]} $ in such a way that it minimises $E$. This optimisation problem is solved sequentially for all tensors in the tree.

\section{Confinement in three dimensional compact QED}

\begin{figure}[t!]
\includegraphics[width=\linewidth]{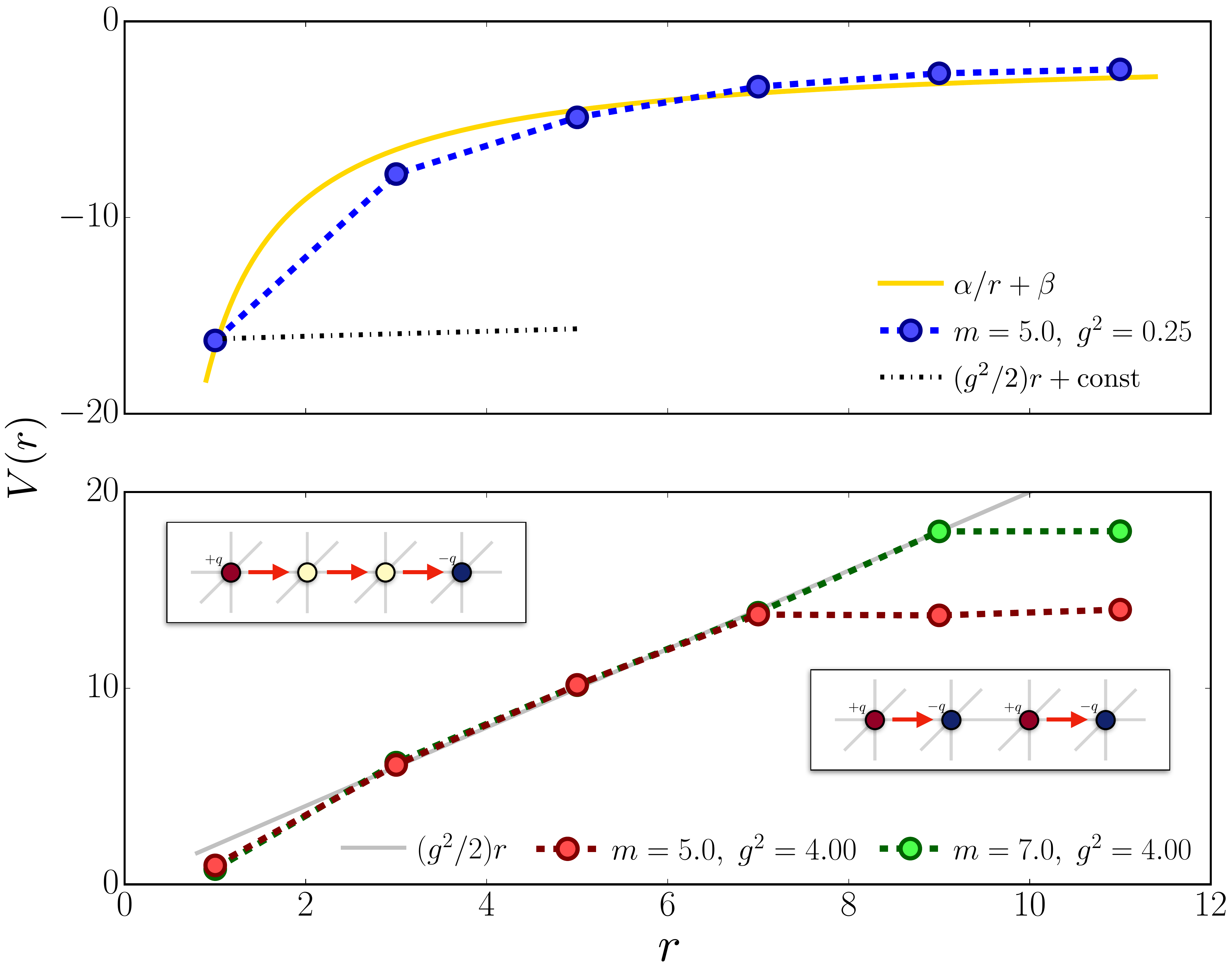}
\caption{ \label{fig:potential}
{Weak coupling regime $g\ll1$ displaying a Coulomb-like potential between two static charges (upper panel) in contrast in the strong coupling regime $g\gg1$ a linear scaling potential is clearly visible (lower panel). The string-breaking radius where the potential becomes distance-independent depends on the different values bare mass (green and red circles). The insets displays cartoons states of the electric field in the string configuration (upper-left) and broken-string one (lower-right). Figure originally presented in Ref.~\cite{Magnifico_QED3D}, courtesy of the authors.}}
\end{figure}

As an paradigmatic example of the possible analysis that are enabled by TN methods, we report here a result recently presented in Ref.~\cite{Magnifico_QED3D} by some of the authors, where the first analysis of a three dimensional lattice gauge theory in presence of matter -- a compact version of QED -- has been reported. Following the steps introduced in Sec. 3(a), and considering $\hbar = c = 1$, the Hamiltonian in three-spatial dimensions simplifies to
\begin{subequations}
\bea 
& \hat H & = 
- \sum_{ \mathbf{{x}}, {\mathbf{\mu}}}
\left(\hat \psi^{\dag}_{{\mathbf{x}}} \, \hat U_{{\mathbf{x}},{\mathbf{\mu}}} \, \hat \psi_{{\mathbf{x}}+{\mathbf{\mu}}} + \text{H.c.} \right) \label{eq_line: kinetic} \\
&+& m \sum_{{\mathbf{x}}}(-1)^{{\mathbf{x}}} \hat \psi^{\dag}_{{\mathbf{x}}}\hat \psi_{{\mathbf{x}}}  +  \frac{{ g^2}}{2} \sum_{ {\mathbf{x}},{\mathbf{\mu}}} \hat L_{{\mathbf{x}},{\mathbf{\mu}}}^{2} \label{eq_line: electric_and_mass} \\
&-& \frac{{ 4}}{g^2}\sum_{x}  \left( \hat \square_{{\mathbf{\mu}_x}, {\mathbf{\mu}_y}}  + \hat \square_{{\mathbf{\mu}_x}, {\mathbf{\mu}_z}} +  \hat \square_{{\mathbf{\mu}_y}, {\mathbf{\mu}_z}} + \text{H.c.} \right) \label{eq_line: magnetic_plaquette}
\eea
\end{subequations}
\\
where ${\mathbf{x}}  \equiv \left ( i,j,k \right )$, $0 \leq i,j,k \leq  L-1$, $ \hat \square_{{ \mathbf{\mu}_\alpha}, { \mathbf{\mu}_\beta}} = \hat U_{{\mathbf{x}},{ \mathbf{\mu}_{\alpha}}}  \hat U_{{\mathbf{x}}+{ \mathbf{\mu}_{\alpha}},{\mathbf{\mu}_\beta}} \hat U^{\dag}_{{ \mathbf{x}}+{ \mathbf{\mu}_\beta},{\mathbf{\mu}_\alpha}}  \hat U^{\dag}_{{ \mathbf{x}},{\mathbf{\mu}_\beta}}$, and we set $\hbar=c=a=1$.  

Fig. \ref{fig:potential} reports a numerical simulation reporting the potential between two charges as a function of the distance $V(r)$ for different bare masses and $m$ and coupling $g$, showing the confinement crossover~\cite{WilsonConfinement, POLYAKOV197582, BANKS1977493}.  The simulation has been performed on a cylinder of dimension $16 \times 4 \times 4$ lattice sites, where two opposite charges are pinned via a strong local chemical potential at distance $r=1, 2, \dots, 11$. In weak coupling $g^2 \ll 1$, the potential$V(r)$ is Coulomb like: the difference between the ground state energy  $E$ of the system with additional charges and without $E_0$ scales as $V(r)= E -E_0 \sim 1/r$~\cite{Guth80}. On the contrary, in the strong coupling limit $g^2 \gg 1 $, the magnetic term in Eq.~\eqref{eq_line: magnetic_plaquette} becomes negligible and it is more energetically favourable for the field lines to be confined in a string connecting the two static charges, as witnessed by the potential between the charges that becomes linear in the distance, $V(r) \propto r$. The latter picture holds until the energetic cost of creating a particle-antiparticle pair from vacuum is less that the string potential: depending on the particle bare mass $m$, there is a critical string length beyond which the string breaks and the charges can be separated at no extra energy cost. This genuinely three dimensional quantum behaviour is confirmed by the TN numerical simulations displayed in Fig. \ref{fig:potential}. 

Finally, we mention that in~\cite{Magnifico_QED3D} an extensive analysis of the model has been reported, including screening effect in a quantum capacitor and the characterisation of its ground state properties at zero and finite global chemical potential. We stress that the latter simulations (not reported here) efficiently attack a challenge inefficient to tackle with Monte Carlo methods due to the sign problem arising in that context. 

\section{Conclusions and perspectives}

In this short review, we have presented the brief collection of the different concepts on which a quickly expanding field is based on: the tensor network paradigm and some of the most successful tensor network ansatz states and algorithms used to perform numerical characterisation of different phenomena. Their application to the study of lattice gauge theories requires to recast the theories in the Hamiltonian formulation, and the necessity of keeping the algorithms as efficient as possible calls for additional technical but fundamental theoretical approaches, such as the introduction of the rishon representation of the gauge degrees of freedom that allows to recast the gauge constraint in an easy-to-satisfy redefinition of the local basis plus an ad-hoc additional Abelian gauge constraint. The advantage of the rishon representation of the gauge degree of freedom is that the only remaining symmetry is always Abelian, independently from the nature -- Abelian or non-Abelian -- of the original theory of interest, enormously simplifying the technical implementation of symmetric ansatz states. Moreover, the rishon representation allows to defermionize the theory, that is, to rewrite the Hamiltonian in terms of only bosonic operators, simply and elegantly solving the highly technically complex implementation of the fermionic algebra of matter fields, e.g., without the need of introducing long-range interactions (Jordan-Wigner strings).   

The scope of this brief overview is to introduce the interested reader to the main concepts and to present in a clear and concise form the different steps that altogether are opening the possibility of studying lattice gauge theories in regimes where Monte Carlo simulations are inefficient. We are aware that these pages are only scratching the tip of the iceberg of the long list of powerful techniques that has been introduced in the last decades and that it is constantly improving: we invite the interested reader to approach longer and comprehensive reviews and texts presenting tensor network methods in general~\cite{RomanReview2019,Simone2019BookReview,TNAnthology,MCB2020review} and their application to lattice gauge theories, including -- to name a few -- results for non-Abelian theories~\cite{MCBNonAbelian2015, FerdinandSU2,YannickSU3}, time-dependent and scattering problems~\cite{ThomasStringBreak}, and systems at finite temperature~\cite{3DFiniteTLGT}.

Finally, we would like to highlight three different but interconnected research directions that are explored by an increasing interdisciplinary community that are opening new exciting possibilities. Indeed, tensor networks are currently the most versatile and powerful tool to simulate quantum many-body systems and in particular quantum computers and quantum simulators. They are currently used as a benchmark for some of the cutting edge quantum computation and quantum simulations and are setting the quantum advantage threshold~\cite{QSupremacyNeven2018}. However, digital and analogue quantum computations and quantum simulations have also been proposed for studies of lattice gauge theories of increasing complexity on different platforms~\cite{Martinez2016QSim,IonsLGTQSimKokail,AtomsLGTQsimTagliacozzo,wiese2013ultracold,AtomsLGTQsimZohar,SupercondLGTQsimMarcos,RydbergLGTQsimSurace,kan2021, Monika2019floquet, Banuls2019, mil2020scalable, yang2020observation}, implementing or proposing paths to implement Feynman's seminal idea on NISQ era devices~\cite{FeynmanQSim82,PreskillQSim2018}. The interplay between tensor network simulations and real quantum computation and simulation will likely lead the development of the field in the next years \cite{klco2021QsimRev,zohar2021QsimRev}.

Similarly, the very same tools can be applied also to condensed matter systems and beyond, such as quantum chemistry and material science. Interestingly, in the condensed matter there is a large class of systems that are lattice gauge theories, describing low-lying energy properties of complex materials or paradigmatic models displaying topological systems, e.g., quantum spin ice and the Kitaev model~\cite{FerdiSpinIce,ToricCode,GersterFQH2017,TopoTNHierarchy}.  The tools and know-how developed in both fields can be used to foster the development of the other. Moreover, hard optimisation problems in computer science can be recast in disordered spin lattice models which, in turn, can be mapped in two-dimensional lattice gauge theories via the LHZ mapping~\cite{LHZ}. Thus, novel insights on computer science problems can be obtained via this elegant detour.  

The last cross-fertilisation that we are witnessing among different fields is that between machine learning and quantum science. The development of quantum machine learning is well established by now~\cite{MachineLearning2015,Biamonte2016,Liu2020b}, however, recently it has been recognised that tensor networks can be used to perform machine learning tasks such as supervised and unsupervised learning~\cite{AIRussell02,AIMohri18}. As, for quantum simulation and computations, this approach is a subset of the possible quantum machine learning tools, those quantum circuits that can be efficiently simulated classically. However, despite not being fully general, this quantum-inspired machine learning  (or tensor network machine learning) has the advantage of overcoming some of the challenges of quantum machine learning (i.e., the loading of big data, the learning process, etc.) and has already proven to be in the tested cases as effective as deep neural learning~\cite{Stoudenmire2016a,Huggins2019}. In particular, this approach has been recently applied to HEP related problems: the classification of $b$-$\bar b$ asymmetry in the LHCb experiment showing interesting results and new possible development directions~\cite{Zuliani2021a}. Other quantum machine learning and more generally quantum algorithms for HEP problems are explored with increasing interests to attack the future challenges of the field, especially those posed by the future high-luminosity LHC runs~\cite{HiLumiLHC,Tuysuz2020,Guan2021}. 

In conclusion, we hope that this brief introduction to the field has triggered the interests of the reader as we believe that tensor network methods, and more generally the interplay between quantum science and high-energy physics~\cite{klco2021standard}, will lead to new fascinating opportunities and exciting developments in the next years. 

{\emph{Acknowledgments -}} This work is partially supported by the Italian PRIN 2017, by fondazione CARIPARO, the INFN project QUANTUM, the European Union Horizon 2020 research and innovation programme under grant agreements No. 817482 (PASQuanS), and the  QuantERA QTFLAG and QuantHEP projects, and the DFG project TWITTER.

\competing{The authors declare that they have no competing interests.}
\footnotesize{
\bibliographystyle{vancouver}
\bibliography{Refs}
}
\end{document}